\documentclass[]{eptcs}

 \providecommand{\event}{QFM09}
 \providecommand{\volume}{}
 \providecommand{\anno}{2009}
 \providecommand{\firstpage}{1}

\usepackage[fleqn]{amsmath}
\usepackage{amssymb}
\usepackage{amscd}
\usepackage{stmaryrd}

\usepackage[all]{xy}\newdir{...}{{}*{}*{}}

\usepackage[amsmath,thmmarks]{ntheorem}
 \theorembodyfont{\upshape}
 \theoremheaderfont{\bfseries}
  \theoremsymbol{\ensuremath{\square}}
  \newtheorem{theorem}{Theorem}[section]
  \newtheorem{definition}[theorem]{Definition}

  \theoremstyle{nonumberplain}
  \theoremsymbol{\ensuremath{\blacksquare}}
  \newtheorem{proof}{Proof}

\newcommand{\bbbr}{\mathbb{R}}
\newcommand{\realn}{\bbbr}

\newcommand{\mspc}[2]{\realn^{#1\times #2}}

\newcommand{\fml}[1]{\ar@(r,u)@{...}^{{#1\!\!}}}
\newcommand{\fmlleft}[1]{\ar@(l,u)@{...}_{#1}}
\newcommand{\termst}[1]{\ar@(ld,rd)@{...}^{\downarrow}}
\newcommand{\initst}[1]{\ar@(lu,ru)@{...}_{\downarrow\vspace{-2mm}}}

\newcommand{\mc}{Markov chain}
\newcommand{\mcs}{Markov chains}
\newcommand{\mrc}{Markov reward chain}
\newcommand{\mrcs}{Markov reward chains}

\newcommand{\rwd}[1]{\ar@(r,u)@{...}^{{#1\!\!}}}
\newcommand{\invec}[1]{\ar@(l,u)@{...}_{#1}}
\newcommand{\inveczero}{}

\newcommand{\ql}{Q_{\!s}}
\newcommand{\qtau}{Q_{\!f}}

\newcommand{\asf}{\mathsf{A}}
\newcommand{\PiV}{\Pi^{(V)}}
\newcommand{\qtauV}{\qtau^{(V)}}
\newcommand{\ab}{\mathbb{P}(\asf)}

\newcommand{\zo}{{\{0,1\}}}
\newcommand{\wmat}{\transp{V}}

\newcommand{\vone}[1]{\mathbf{1}^{#1}}
\newcommand{\vzero}[1]{\mathbf{0}^{#1}}
\newcommand{\transp}[1]{\mathalpha{{#1}^\mathsf{\tiny T}}}

\newcommand{\goes}[1]{\ensuremath{\mathbin{\stackrel{#1}{\rightarrow}}}}

\newcommand{\term}[1]{{#1}\mathalpha{\downarrow}}
\newcommand{\nterm}[1]{{#1}\mathalpha{\not\downarrow}}

\newcommand{\mtsna}[2]{\mathcal{T}_{#2}^{#1}}
\newcommand{\mts}{\mtsna{n}{\asf}}

\newcommand{\triple}[3]{\mathalpha{\langle#1,#2,#3\rangle}}
\newcommand{\quadriple}[4]{\mathalpha{\langle#1,#2,#3,#4\rangle}}

\newenvironment{mypmatrix}{\setlength{\arraycolsep}{0.6\arraycolsep}\begin{pmatrix}}
  {\end{pmatrix}}

\title{Strong, Weak and Branching Bisimulation for Transition Systems and Markov Reward Chains: A Unifying Matrix
Approach}

\author{Nikola Tr\v{c}ka
  \institute{
  Department of Mathematics and Computer Science,\\
  Eindhoven University of Technology,\\
  P.O.\ Box 513, NL-5600 MB Eindhoven, The Netherlands
  }
  \email{n.trcka@tue.nl}
  }

\def\titlerunning{Strong and Weak Bsm for LTSs and MRCs}
\def\authorrunning{N. Tr\v{c}ka}

\begin{document}
\maketitle
\begin{abstract}
We first study labeled transition systems with explicit successful termination.
We establish the notions of strong, weak, and branching bisimulation in terms
of boolean matrix theory, introducing thus a novel and powerful algebraic
apparatus. Next we consider Markov reward chains which are standardly presented
in real matrix theory. By interpreting the obtained matrix conditions for
bisimulations in this setting, we automatically obtain the definitions of
strong, weak, and branching bisimulation for Markov reward chains. The obtained
strong and weak bisimulations are shown to coincide with some existing notions,
while the obtained branching bisimulation is new, but its usefulness is
questionable.
\end{abstract}


\section{Introduction}
(Labeled) transition systems are a well established formalism for modeling of
the qualitative aspects of systems, focusing on the behavioral part. A
transition system is a directed graph in which nodes represent states of the
system, and labels on arrows represent the actions that the system can perform
when going from one state to another. One state is considered to be the
starting state. It is sometimes useful to distinguish deadlock behavior
(inability to proceed) from successful termination, so we consider transition
systems in which some states are explicitly marked as (successfully)
terminating~\cite{emptyprocess,empty-baeten-time}.

Reasoning about transition systems is usually done by relating them according
to some behavioral equivalence. If two systems are to agree on every step they
take, then they are equivalent modulo \emph{strong bisimulation}
equivalence~\cite{park,milner}. When a system can perform internal (silent)
steps, of which the impact is considered unobservable, strong bisimulation is
less appropriate because it equates too few states. To solve this problem
weaker equivalences have been introduced that abstract away from the internal
steps but require that the other, i.e.\ visible, steps are strongly simulated.
The two most commonly used equivalences of this type are \emph{weak}
bisimulation~\cite{milner} and \emph{branching} bisimulation
\cite{branching,basten} equivalence. The difference between the two is that the
latter preserves the branching structure of a transition system
better~\cite{branching}.

While transition systems are very useful for qualitative reasoning,
(continuous-time) \mcs{}  have established themselves as powerful, yet fairly
simple models for performance evaluation, i.e., for the modeling of
quantitative behavior of systems. A \mc{} can be represented as a directed
graph in which nodes represent states and labels on the outgoing arrows
determine the stochastic behavior (an exponentially distributed delay) in the
state. Some states are marked as starting and have initial probabilities
associated with them. To increase modeling capability and obtain some very
useful performance measures, such as throughput and utilization of a system,
\mcs{} are often equipped with \emph{rewards}~\cite{howard}. We only consider
rewards associated to states, in which case they actually represent the rate
with which a \mc{} gains a reward while residing in a state. A \mc{} with
(state) rewards is called a \mrc{}.

The idea of strong bisimulation exists in the \mrc{} world as an aggregation
method called \emph{ordinary lumping} \cite{kemeny,nicola,bucholtz}.\footnote{
To be fair we could also say that strong bisimulation is the transition system
analogue of ordinary lumping.} This method is based on joining states that have
the same reward and that transfer to lumping classes  with equal probabilities
at any given time. This ensures that the stochastic behavior of strongly
bisimilar states is the same, and that bisimulation keeps (the reward based)
performance properties. The notions of weak and branching bisimulation,
however, have not yet been introduced to pure \mcs{} (one exception is the weak
bisimulation of~\cite{baier1} that actually coincides with ordinary lumping)
but only to their  extensions  coming from stochastic process
algebras~\cite{HermannsPhD}. All these extensions add action information to
\mcs{}, and, like in transition systems, take the special action $\tau$ for the
internal step that can be abstracted from. The abstraction is not
stochastically formalized, in the sense that it is not defined what performance
properties are shared among weakly bisimilar chains. Moreover, when restricted
to exponential transitions, the existing equivalences typically coincide with strong bisimulation. 


In this paper we establish the notion of strong, weak, and branching
bisimulation for \mcs{} by setting the theory of transition systems and \mcs{}
on a common ground. The well developed matrix apparatus has shown to be a
powerful method for presenting and reasoning about \mcs{}, so we take matrix
theory for the unifying framework. The three notions of bisimulation on
transition systems are first formalized in (boolean) matrix terms, leading to a
system of matrix equalities. These equalities can be directly interpreted in
the (standard matrix) setting of \mcs{}, automatically yielding  definitions of
strong, weak, and branching bisimulation there. The obtained notion of strong
bisimulation is proven to indeed coincide with the standard definition of
ordinary lumping. The obtained notion of weak bisimulation is shown to
remarkably coincide with the notion of $\tau$-lumping that we have recently
developed as a helping tool in solving a different and independent
problem~\cite{QEST2006,TrckaPhD}. To the best of our knowledge, the obtained
definition of branching bisimulation does not correspond to any known Markovian
equivalence from the literature.

With the decision to use matrix theory as a common setting, the gain is
twofold. The matrix approach to transition systems sets the theory in a
powerful algebraic setting  that can be used as an alternative to or in
combination with the standard process algebraic approach. This is specially
useful  because the notion of (bi)simulation has been, in some forms,
extensively studied in graph, modal logic, and automata theory~\cite{schmidt,fitting,buchholz08}.
Matrices, moreover, increase clarity and compactness, simplify proofs, make
known results from linear algebra directly applicable, have didactic advantage,
etc.


The structure of this paper is as follows. In the next section we give some
preliminaries for working with matrices. In Section~\ref{sec::lts} we define
 transition systems with explicit termination as systems of matrices. We
also give matrix definitions of strong, weak, and branching bisimulation, and
show that these notions indeed correspond to the standard ones. In
Section~\ref{sec::mc} we first standardly present the theory of \mrcs{} using
matrix theory. Then we interpret the definitions from Section~\ref{sec::lts} in
this setting and discuss the resulting notions of bisimulation. The last
section gives some conclusions and directions for future work. 

\section{Preliminaries}
Let $X$ be a set with addition, multiplication, the unit elements $0$ and $1$
for these operations, and with a preorder $\leq$. Then $X^{n\times m}$ denotes
the set of all $n\times m$ matrices with elements in $X$. We assume that matrix
addition, matrix multiplication, multiplication by a scalar, and $\leq$, are
all standardly defined in $X^{n\times m}$. The \emph{elementwise product} of
two matrices is defined as $(A\sqcap B)[i,j] = A[i,j]B[i,j]$.

Elements of $X^{1\times n}$ and $X^{n\times 1}$ are called (row and column)
vectors. $\vone{n}$ denotes the vector in $X^{n\times 1}$ that consists of $n$
$1$'s. $\vzero{n\times m}$ denotes the $n\times m$ matrix consisting entirely
of zeroes. $I^n$ denotes the $n\times n$ \emph{identity matrix}. We omit the
$n$ and $m$ when they are clear from the context. A matrix $A$ of which every
element is either $0$ or $1$, i.e.\ an element of $\zo^{n\times m}$, is called
a $0$--$1$ matrix.


A $0$--$1$ matrix $V\in X^{n\times N}$, $n\geq N$ in which every row contains
exactly one $1$ is called a \emph{collector}. For the theory of bisimulation
the central notion is of partitioning of states into equivalence classes. We
can then think of a collector matrix as a matrix in which the rows represent
states, the columns represent the equivalence classes, and the entries indicate
 which states belong to which classes. Note that
 $V\cdot\vone{} =\vone{}$. A matrix $U\in X^{N\times n}$
such that $U\cdot \vone{} = \vone{}$ and $UV=I^N$ is a \emph{distributor} for
$V$.

\section{Transition Systems in Matrix Terms}\label{sec::lts}
Let $\asf$ be a set and let $\mathcal{P}(\asf)$ be the set of all subsets of
$\asf$. Then $\ab=(\mathcal{P}(\asf),+,\cdot,\bar{\phantom{a}},0,1)$ is a boolean
algebra with $+$, $\cdot$, $\bar{\phantom{a}}$, $0$ and $1$ representing union,
intersection, complement, the empty set and the full set $\asf$ respectively.
We use $+$, $\cdot$, $0$ and $1$ instead of $\cup$, $\cap$, $\emptyset$ and
$\asf$ to emphasize the connections with standard matrix theory.

We now assume that $\asf$ is a set of actions and fix it for the reminder of
this section. A transition system is standardly defined as a quadruple
$(S,\rightarrow,S_0,\downarrow)$ where $S$ is a finite set of states,
${\rightarrow} \subseteq S \times \asf\times S$ is the \emph{transition
relation}, $s_0\in S$ is the \emph{initial state} and $\downarrow \subseteq S$
is the set of \emph{(successfully) terminating states}.

In matrix terms we define a transition system as a triple of a  $0$--$1$ row
vector that indicates which of the states is initial, a matrix whose elements are sets
of actions that the system performs when transiting from one state to another, and a $0$--$1$  vector that indicates which states are terminating.

\begin{definition}[Transition system]
    A \emph{transition system\/} (of the dimension $n$) is a triple $\triple{\sigma}{A}{\rho}$ where:
    \begin{itemize}
    \item $\sigma\in \zo^{1 \times n}$ is the \emph{initial vector} with exactly one non-zero entry,
    \item $A\in\ab^{n\times n}$ is  the \emph{transition matrix}, \quad and
    \item $\rho\in \zo^{n\times 1}$  is the \emph{termination vector}.
    \end{itemize}
     The set of all transition systems
       of the dimension $n$ is denoted $\mts$.
\end{definition}

\begin{figure}[h!]
\begin{displaymath}\small
\begin{array}{cccc}
\begin{array}{c}
\xymatrix@R=0.9cm@C=0.3cm{
  & *++[o][F] {s_1}\initst{}\termst{} \ar[ld]_{a} \ar[rd]^{a}\\
        *++[o][F] {s_2} \ar@/_/[rd]_{b} \ar@/^/[rd]^{c}  & & *++[o][F] {s_3} \ar@/_/[ld]_{b}\\
        &  *++[o][F]{s_4}\termst{} \ar@/_/[ru]_{d} & }
\end{array}
  &&&
    \begin{array}{l}
    \sigma=\begin{mypmatrix} 1 & 0 & 0 & 0\end{mypmatrix}\\\\
        A=\begin{mypmatrix} 0 & \{a\} & \{a\} & 0 \\
        0 & 0 & 0 & \{b,c\} \\
        0 & 0 & 0 & \{b\} \\
        0 & 0 & \{d\} & 0
        \end{mypmatrix}
        \quad
        \rho= \begin{mypmatrix} 1 \\
        0 \\
        0 \\
        1
        \end{mypmatrix}.
    \end{array}
\end{array}
\end{displaymath}
 \caption{Transition system and its matrix representation}
 \label{fig::tran-sys-examp}
\end{figure}
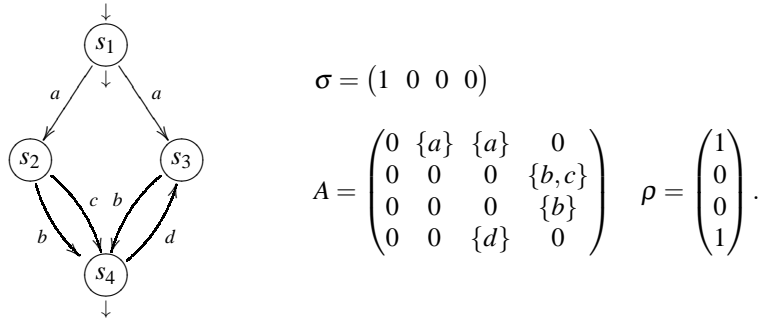

If $S=\{s_1,\ldots,s_n\}$, our definition is obtained from the standard one by
putting:
\[ A[i,j]= \{a \mid s_i\goes{a} s_j\}, \ \
\sigma[i]=\left\{\begin{array}{rl} 1, & \text{if } s_i = s_0 \\ 0, & \text{if }
s_i \neq s_0
\end{array}\right.  \text{\ \  and \  } \rho[i]=\left\{\begin{array}{cl} 1,
& \text{if }\term{s_i} \\ 0, & \text{if }\nterm{s_i}.
\end{array}\right.\]
That is, for each two states $s_i$ and $s_j$, $A[i,j]$ contains the set of
actions that the system can perform by going from $s_i$ to $s_j$. The $i$-th
element of $\sigma$ is $1$ if the state $s_i$ is initial. The $i$-th element of
$\rho$ is either $0$ or $1$ depending if the state $s_i$ is terminating or not.
It is  clear that, given an ordered set $S$, we can obtain the standard
definition from our definition easily. Figure~\ref{fig::tran-sys-examp} depicts
a transition system and gives its matrix representation. The set of states is
$S=\{s_1,s_2,s_3,s_4\}$. State $s_1$ is the initial state; states $s_1$ and
$s_4$ are terminating.

\subsection{Strong bisimulation}
Strong bisimulation is an equivalence relation that partitions the set of
states in such a way that the set of actions that can be executed to reach some
class is the same for every two states in a class. In addition, the termination
behavior of two related states must be the same. This allows us to built the
quotient (i.e.\ the lumped) system in which states are the equivalence classes.
In matrix terms bisimulation  conditions and the lumped system are conveniently
expressible in terms of a collector and a distributor matrix.

\begin{definition}[Strong bisimulation-LTS]\label{def::str-bsm-lts}
A collector matrix $V\in\zo^{n\times N}$ is called a \emph{strong bisimulation
on} the transition system $\triple{\sigma}{A}{\rho}\in\mts$ if
\begin{equation*}
    VUAV = AV \text{ and } VU\rho = \rho,
\end{equation*}
where $U$ is some distributor for $V$. In this case we also say that
$\triple{\sigma}{A}{\rho}$ \emph{strongly lumps (by $V$)} to the transition
system $\triple{\hat{\sigma}}{\hat{A}}{\hat{\rho}}\in\mtsna{N}{\asf}$ defined
by:
    \[\hat{\sigma} = \sigma V,\quad\hat{A}= U  A V \quad\text{and}\quad
    \hat{\rho}= U \rho.\]
\end{definition}

\noindent Both the conditions for bisimulation and the definition of the lumped
process do not depend on the particular choice of a distributor. Suppose that
$U' V = I$ for some $U'\neq U$. Then $VU'A V = V U' V U A V = V U A V = A V$
and similarly $VU\rho = \rho$. Also $U'A V = U' V U A V = U A V$ and similarly
$U'\rho = U\rho$.

We now show that our definition of strong bisimulation agrees with the standard
one. Define $R= V\transp{V}$ and note that $\transp{V}$ is a also distributor for $V$. Clearly $R\geq I$. We first prove that the above conditions are equivalent to the conditions $RA\leq AR$ and $R\rho\leq\rho$. We calculate $RA = V\transp{V}A \leq  V\transp{V}A V\transp{V} =A V\transp{V} = AR$ and $R\rho =  V\transp{V}\rho \leq  \rho$, and for the other direction, $AV\leq V\transp{V} A V \leq A  V\transp{V} V = AV$ and $\rho \leq
V\transp{V}\rho \leq \rho$.  Note now that $a\in (RA)[i,j]$ iff there is a $k$
such that $R[i,k]=1$ and $a\in A[k,j]$. Similarly, $a\in (AR)[i,j]$ iff there
is an $\ell$ such that $a\in A[i,\ell]$ and $R[\ell,j]=1$. The condition $R A
\leq A R$ then says that
\[\begin{array}{c}\xymatrix@R0.7cm@C1cm{s_i\ar@{--}[r]^{R} & s_k\ar[d]_{a} \\ & s_j}\end{array} \qquad \text{implies}\qquad
    \begin{array}{c}\xymatrix@R0.7cm@C1cm{ s_i \ar[d]^{a}&   \\ s_\ell\ar@{--}[r]_{R} &
s_j.}\end{array}
\]
This clearly corresponds to the standard definition of strong bisimulation.
Finally, note that $(R\rho)[i]=1$ iff there is a $j$ such that $R[i,j]=1$ and
$\rho[j]=1$. Thus, the condition $R\rho\leq \rho$ says that:
\[\begin{array}{c}\xymatrix@R0.1cm@C1cm{s_i \ar@{--}[r]^{R}& s_j\term{}}\end{array} \qquad \text{implies}\qquad \begin{array}{c}\xymatrix@R0.5cm@C1cm{\term{s_i}.}\end{array}
\]
This again matches with the standard definition.

\subsection{Weak bisimulation}

A silent step in a transition system is a step that is labeled by the internal
action $\tau$. Every matrix $T\in\ab^{n\times n}$ can be uniquely represented
as $T = A + \{\tau\}\cdot S$ where $\tau\in\asf$, and $A,S\in\ab^{n\times n}$
are such that $\{\tau\}\cdot A = \vzero{}$ and $S$ is a $0$--$1$ matrix. To
make this form of $T$ more explicit we write $\quadriple{\sigma}{A}{S}{\rho}$
instead of $\triple{\sigma}{T}{\rho}$. Note that the strong bisimulation
conditions  from the previous section can be decomposed into separate
conditions on $A$ and $S$. In other words, the condition $VUTV = TV$ is valid
if and only if the inequalities $VUAV = AV$ and $VUSV = SV$ both hold.

Weak bisimulation~\cite{milner}  ignores silent transitions in a very general
way. It requests that a transition labeled with an action is simulated by a
transition labeled with the same action but preceded and followed by a sequence
of $\tau$ transitions. For this we introduce a matrix definition of
reflexive-transitive closure. Given a $0$--$1$ matrix $R\in\zo^{n\times n}$, we
call the matrix $R^* =  \sum_{n=0}^\infty R^n$ the \emph{reflexive-transitive
closure} of $R$.

\begin{definition}[Weak bisimulation]\label{def::weak-bsm-lts}
A collector matrix $V\in\zo^{n\times N}$ is a \emph{weak bisimulation} on the
transition system $\quadriple{\sigma}{A}{S}{\rho}\in\mts$ iff
\[VU\Pi V = \Pi V,  VU\Pi A \Pi V = \Pi V, \text{ and } VU\Pi\rho = \Pi\rho\] where
where $\Pi=S^*$ is the reflexive transitive closure of $S$ and $U$ is some
distributor for $V$. We say that $\quadriple{\sigma}{A}{S}{\rho}$ \emph{weakly
lumps (by $V$) to}
    $\quadriple{\hat{\sigma}}{\hat{A}}{\hat{S}}{\hat{\rho}}$ defined by
    \[\hat{\sigma} = \sigma V,\quad\hat{A}= \transp{V}  A V \quad\text{and}\quad
    \hat{\rho}= \transp{V} \rho.\]
\end{definition}

\noindent Contrary to strong bisimulation the definition of the lumped
    process now depends on the distributor used. Using  any other
distributor would, in general, give a different result for the lumped system
(the irrelevance of distributors is only implied by the \emph{strong} bisimulation condition, i.e. by $VUAV = AV$ and $VU\rho = \rho$, which might not hold here).

Our definition of weak bisimulation corresponds to the standard one.  First,
$S^*[i,j]=1$ iff there is an $n\geq 0$ such that $S^n[i,j]=1$. This is
equivalent to saying that there exist $i_0,\ldots, i_n$ such that $i_0=i$,
$i_n=j$ and $S[i_k,i_{k+1}]=1$ for all $k=0,\ldots, n-1$. Recall that
$S[i,j]=1$ means, in the standard theory, that $s_i\goes{\tau} s_j$. Thus,
$\Pi[i,j]=S^*[i,j]=1$ means that we have $s_{i_0}\goes{\tau}\ldots\goes{\tau}
s_{i_n}$ or that, in the standard notation, $s_i\Rightarrow s_j$. Now, as we
did for strong bisimulation, we let $R=V\transp{V}$ and express the
bisimulation conditions using $R$. We have $RS \leq R\Pi \leq R\Pi R =
V\transp{V} \Pi V = \Pi V\transp{V} = \Pi R$, $RA \leq R \Pi A \Pi R =
V\transp{V} \Pi  A
 \Pi V\transp{V} = \Pi  A \Pi V\transp{V} = \Pi  A \Pi R$, and  $R\rho \leq R\Pi\rho \leq \Pi \rho$.
The first inequality means that
\[\small\begin{array}{c}
    \xymatrix@R0.5cm@C1cm{s_i\ar@{--}[r]^{R}  & s_k\ar@{->}[d]^{\tau}\\
                                                    & s_j}
  \end{array} \qquad \text{implies}\qquad
  \begin{array}{c}\xymatrix@R0.5cm@C1cm{s_i\ar@{=>}[d] & \\
                            s_\ell\ar@{--}[r]^{R} & s_j.}
  \end{array}
\]
For the second inequality note that $a\in (RA)[i,j]$ iff there is a $k$ such
that $R[i,k]=1$ and $a\in A[k,j]$. Now, $a\in (\Pi A \Pi R)[i,j]$ iff there
exist $1\leq \ell,\ell',\ell''\leq n$ such that $\Pi[i,\ell']=1$, $a\in
A[\ell',\ell'']$, $\Pi[\ell'',\ell]=1$ and $R[\ell,j]=1$. Therefore, $RA\leq
\Pi A \Pi R$ means that
\[\small\begin{array}{c}
    \xymatrix@R1.3cm@C1cm{s_i\ar@{--}[r]^{R}  & s_k\ar@{->}[d]^{a}\\
                                                    & s_j}
  \end{array} \qquad \text{implies}\qquad
  \begin{array}{c}\xymatrix@R0.4cm@C1cm{s_i\ar@{=>}[d] & \\
                             \ar@{->}[d]^{a} & \\
                             \ar@{=>}[d] & \\
                            s_\ell\ar@{--}[r]^{R} & s_j,}
  \end{array}
\]
for $a\neq\tau$. Finally,  $R\rho\leq \Pi\rho$ means that
\[\small\begin{array}{c}\xymatrix@R0.1cm@C1cm{s_i \ar@{--}[r]^{R}& \term{s_j}}\end{array} \qquad \text{implies}\qquad
      \begin{array}{c}\xymatrix@R0.5cm@C1cm{s_i\ar@{=>}[d] & s_j.\\
                            \term{s_\ell}\ar@{--}[ur]_{R} & }
  \end{array}
\]
This is the standard definition of weak bisimulation (with explicit
termination).

Weak bisimulation can also be interpreted as a strong bisimulation on a system
closed under the sequence of $\tau$ transitions, inducing the following
diagram:
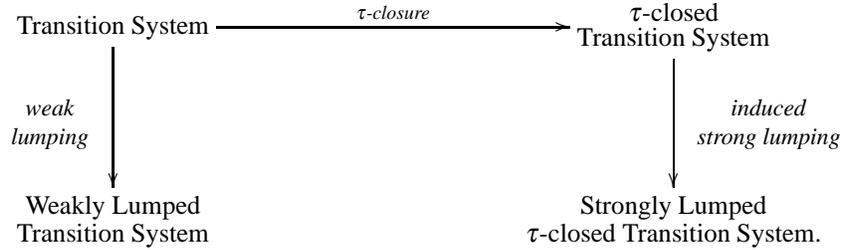
\begin{displaymath}\small
 \xymatrix@R1.7cm@C4cm{
 \txt{Transition System} \ar[r]^{\emph{$\tau$-closure}} \ar[d]_{\footnotesize\begin{tabular}{c}\emph{weak}\\
    \emph{lumping}\end{tabular}} &  \txt{$\tau$-closed\\ Transition System}
                                        \ar[d]^{\footnotesize\begin{tabular}{c}
                                                        \emph{induced}\\
                                                        \emph{strong lumping}
                                                        \end{tabular}} \\
 \txt{Weakly Lumped\\ Transition System}   & \txt{Strongly Lumped \\$\tau$-closed Transition System.}
 }
\end{displaymath}


\noindent We show that weak lumping is sound in the sense that also
\[\small
\xymatrix@R2cm@C4cm{
 \txt{Weakly Lumped\\ Transition System}  \ar[r]^{\emph{$\tau$-closure}} & \txt{Strongly Lumped \\$\tau$-closed Transition System.}
 }\]
The main purpose of the proof is to illustrate the power of matrices in this
setting. We first prove two important properties of $\transp{V}$.

\begin{theorem}\label{thm::pivw-lts}
$\transp{V}\Pi V = (\transp{V}SV)^*$ and $\Pi  V \transp{V} = \Pi V \transp{V}
 \Pi$.
\end{theorem}
\begin{proof}
 \sloppy{We have  $\transp{V}\Pi V = \transp{V} \left(\sum_{n\geq 0} S\right) V \leq
\transp{V} \left(\sum_{n\geq 0} SV\transp{V}\right) V  = \sum_{n\geq 0}
\transp{V}SV = (\transp{V}SV)^*$ and  $\Pi V \transp{V} \leq \Pi V \transp{V}
\Pi \leq \Pi V \transp{V} \Pi V
    \transp{V} = \Pi \Pi V \transp{V} = \Pi V \transp{V}$.}
\end{proof}

Using this theorem and the conditions of Definition~\ref{def::weak-bsm-lts} we
calculate  $(\wmat S V)^* \wmat  A V (\wmat SV)^*
                = \wmat \Pi V \wmat  A V \wmat \Pi V
                =  \wmat \Pi V \wmat  A \Pi V
                =  \wmat \Pi V \wmat  \Pi A \Pi V
                =  \wmat  \Pi \Pi A \Pi V
                =  \wmat  \Pi A \Pi V$
    and $\wmat \Pi\rho=   \wmat \Pi\Pi \rho = \wmat \Pi V \wmat  \Pi \rho = \wmat \Pi V \wmat  \rho=(\wmat S V)^* \wmat
    \rho$,
which exactly states that the order of application of $\tau$-closure and
lumping is irrelevant.

\subsection{Branching bisimulation}

Branching bisimulation~\cite{branching} preserves the branching structure of a
system more than weak bisimulation by requiring that after the initial sequence
of $\tau$ steps the resulting state must again be bisimilar to the same state
that the starting state is bisimilar to.

\begin{definition}[Branching bisimulation]\label{def::branch-bsm-lts}
    A collector $V\in\zo^{n\times N}$ is a \emph{branching  bisimulation} on
    $\quadriple{\sigma}{A}{S}{\rho}\in\mts$ iff
    \[VU (I + \PiV S) V = (I + \PiV S) V,
    VU\PiV A V = \PiV A V, \text{ and } VU\PiV\rho = \PiV\rho\] where
where $\PiV = (S\sqcap V\transp{V})^*$,  and $U$ is some (any) distributor for
$V$. We say that $\quadriple{\sigma}{A}{S}{\rho}$ \emph{branching lumps (by
$V$) to}
    $\quadriple{\hat{\sigma}}{\hat{A}}{\hat{S}}{\hat{\rho}}$ defined by
    \[\hat{\sigma} = \sigma V,\quad\hat{A}= \transp{V}  A V \quad\text{and}\quad
    \hat{\rho}= \transp{V} \rho.\]
\end{definition}

\noindent Note that $\Pi_V = (S \sqcap V\transp{V})^* \leq S^* = S^* (I+S)$,
showing the known result that every branching bisimulation equivalence is also
a weak bisimulation.

To show that our definition indeed induces the notion of branching
bisimulation, we again let $R=\transp{V} V$. The conditions of
Definition~\ref{def::branch-bsm-lts} are easily shown to be equivalent to the
conditions $R S \leq R + \PiV S R$, $RA \leq \PiV AR$, and $R\rho\leq \PiV
\rho$, which are in turn 
easily shown to exactly match the standard conditions of branching
bisimulation.

Similarly as we did for weak, we can interpret branching bisimulation as a
strong bisimulation on a system closed under the sequence of $\tau$-transitions
that now connect  states from the same class only (note that the closure then
depends on the bisimulation). This induces the following diagram:
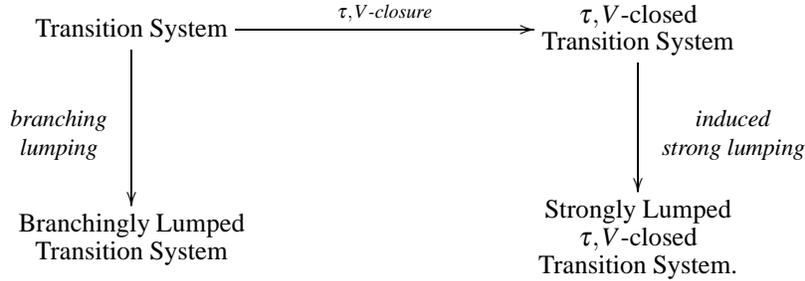
\[\small
\xymatrix@R1.7cm@C3.7cm{
 \txt{Transition System} \ar[r]^{\emph{$\tau, V$-closure}} \ar[d]_{\footnotesize\begin{tabular}{c}\emph{branching}\\
    \emph{lumping}\end{tabular}} &  \txt{$\tau, V$-closed\\ Transition System}
                                        \ar[d]^{\footnotesize\begin{tabular}{c}
                                                        \emph{induced}\\
                                                        \emph{strong lumping}
                                                        \end{tabular}} \\
 \txt{Branchingly Lumped\\ Transition System} & \txt{Strongly Lumped \\$\tau, V$-closed\\ Transition System.}
 }
\]
As it was the case for weak bisimulation, the diagram can be closed, i.e.
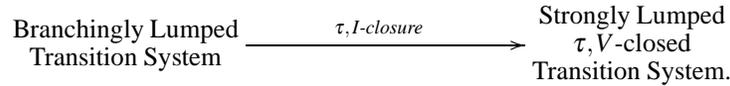
\[\small
\xymatrix@R2cm@C3.7cm{
 \txt{Branchingly Lumped\\ Transition System}  \ar[r]^{\emph{$\tau, I$-closure}}
    & \txt{Strongly Lumped \\$\tau, V$-closed\\ Transition System.}
 }
\]
Since $(\transp{V} SV)^* \sqcap I = I$, this amounts to showing that $I + \wmat
SV = \transp{V} (S^*\sqcap R)(I+S) V$, $\transp{V} AV = \transp{V} (S^*\sqcap
R)A V$ and $\transp{V} \rho = \transp{V} (S^*\sqcap R)\rho$ which  easily
follows from the conditions of
Definition~\ref{def::branch-bsm-lts}.\footnote{We are not aware that this
result has been obtained before, although the corresponding one for weak
bisimulation is known.}

\section{Markov Reward Chains}\label{sec::mc}
We now turn to \mrcs{}. We define the notions of strong, weak and branching bisimulation, by directly interpreting the conditions of the previous section in the real-number matrix setting. For each new notion we discus how it relates to some exiting reduction technique for \mrcs{}.

A \mc{} is a time-homogeneous finite-state stochastic process that satisfies
the Markov property (future independent of the past). It is completely determined by a stochastic \emph{transition matrix (function)} $P(t)$, holding the probabilities of being in some states at time $t>0$, if at a given state at time $0$, and a stochastic row vector that gives the starting probabilities for each state.

The matrix $P(t)$ can conveniently be expressed in terms of a time-independent generator matrix. A \emph{generator} matrix is a square matrix of which
the non-diagonal elements are non-negative and each diagonal element is the
additive inverse of the sum of the non-diagonal elements of the same row. The elements of this matrix are exponential rates It is a standard \mc{} result that for every $P(t)$ there exists a unique generator $Q$ such that $P(t) = e^{Qt}$ (and then also $Q = P'(0)$).

A \mrc{} is a \mc{} where reward is associated to every state, representing the rate at which gain is received while the process is in that state. We now give a formal definition.

\begin{definition}[Markov reward chain]
A \mrc{} is  a triple
$(\sigma,Q,\rho)\in\mspc{1}{n}\times\mspc{n}{n}\times\mspc{n}{1}$ where
$\sigma$ is the \emph{initial probability vector}, $Q$ is a generator matrix
called the \emph{rate matrix}, and $\rho$ is the \emph{reward vector}.
\end{definition}

\begin{figure}[h!]
\begin{displaymath}\small
\begin{array}{cccc}
\begin{array}{c}
\xymatrix@R=0.8cm@C=0.3cm{
& *++[o][F] {s_1} \ar@/_/[ld]_{\lambda} \ar@/^/[rd]^{\mu} \invec{\pi} \rwd{r_1}&\\
*++[o][F]{s_2}\invec{\!\!\!\!\!\!\!\!\!\!\!\!\!\!1\mathord{-}\pi}  &
&*++[o][F]{s_3}\inveczero{} \ar@/^/[ul]^{\nu}\rwd{r_3}}
\end{array}
  &&&
    \begin{array}{l}
    \sigma=\begin{mypmatrix} \pi & 1\mathord{-}\pi & 0 \end{mypmatrix}\\\\
        Q=\begin{mypmatrix} -\lambda\mathord{-} \mu & \lambda & \mu \\
        0 & 0 & 0  \\
        \nu & 0 & -\nu
        \end{mypmatrix}
        \quad
        \rho= \begin{mypmatrix} 
        r_1 \\
        0 \\
        r_3
        \end{mypmatrix}.
    \end{array}
\end{array}
\end{displaymath}
 \caption{A Markov reward chain and its matrix representation}
 \label{fig::mrc-examp}
\end{figure}
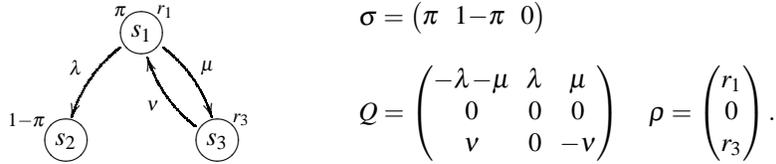

Figure~\ref{fig::mrc-examp} gives an example of a \mrc{}.
This chain starts from state $s_1$ with probability $\pi$ and from state $s_2$
with probability $1-\pi$. In state $s_1$ it waits the amount of time determined by the minimum of
two exponentially distributed delays, one parameterized with rate $\lambda$,
the other with rate $\mu$ (note that this means that the process spends in
state $1$ exponentially distributed time with rate $\lambda + \mu$). After
delaying the process jumps to state $s_2$ or state $s_3$ depending on which of the
two delays was shorter. In state $s_2$ the process just stays forever,
i.e.\ it is absorbed there. From state $s_3$ it can jump back to state $s_1$, after an exponential delay with rate $\nu$. While residing in state $s_i$, for $i=1,3$, the chain earns a reward with rate $r_i$.

If in Definition~\ref{def::str-bsm-lts} of strong bisimulation for transition
systems we replace the transition system with a \mrc{} $(\sigma,Q,\rho)$, we
obtain the following definition of strong bisimulation for \mrcs{}.

\begin{definition}[Strong bisimulation-MRC]\label{def::str-bsm-mrc}
A collector matrix $V\in\zo^{n\times N}$ is called a \emph{strong bisimulation
on} the \mrc{} $\triple{\sigma}{Q}{\rho}$ iff
\[
    VUQV = QV \text{ and } VU\rho = \rho,
\]
where $U$ is some (any) distributor for $V$. In this case we also say that
$\triple{\sigma}{Q}{\rho}$ \emph{strongly lumps} to the \mrc{}
$\triple{\hat{\sigma}}{\hat{Q}}{\hat{\rho}}\in$ defined by:
    \[\hat{\sigma} = \sigma V,\quad\hat{Q}= U  Q V \quad\text{and}\quad
    \hat{\rho}= U \rho.\]
\end{definition}

\noindent While the classical strong bisimulation requires that bisimilar states go to the same equivalence class by performing exactly the same actions, here these states must have equally distributed waiting times and equal joint probabilities when transiting to other classes. Moreover,  they must also have the same reward. The definition reveals the already known fact that strong bisimulation corresponds to the notion of ordinary lumpability for \mrcs{}. The conditions from above exactly match the lumping conditions proposed in
\cite{nicola}. Standard lumping is known to preserve many useful performance properties. For example, the total reward rate at $t$, defined as $R(t) = \sigma P(t) \rho$, is easily shown to be the same for the original and the lumped chain.

We now define the notion of weak bisimulation for \mrcs{}. Similarly to transition systems we  introduce internal steps in a \mrc{} by
assuming that  $Q$ is of the form $\ql + \tau \qtau$, for some (fixed) parameter
$\tau>0$ and two generator matrices $\ql$ and $\qtau$. To indicate this form of
$Q$ we write $(\sigma,\ql,\qtau,\rho)$ for such a \mrc{}.

To be able to apply the classical definition of weak bisimulation 
to \mrcs{} we must first find a matrix $\Pi$ that would correspond to the
notion of reflexive-transitive closure. It is not hard to see that this $\Pi$
must be the \emph{ergodic projection at zero} of $\qtau$, defined by
$\Pi=\lim_{t\rightarrow\infty}e^{\qtau t} = \sum_{n=0}^\infty \qtau^n
\frac{t^n}{n!}$ (the strict formalization of this fact would need to be based
on the theory of eigenvectors for boolean matrices). The matrix $\Pi$ always
exists, and is a stochastic matrix denoting the probabilities that the chain
occupying some state is found in (other) some state in the long run.

Now, as we did for strong bisimulation, putting $(\sigma,\ql,\qtau,\rho)$
instead of $(\sigma,A,S,\rho)$ in Definition~\ref{def::weak-bsm-lts}, we obtain
the following definition of weak bisimulation for \mrcs{}.

\begin{definition}[Weak bisimulation-MRC]\label{def::weak-bsm-mrc}
A collector matrix $V\in\zo^{n\times N}$ is called a \emph{weak bisimulation
on} the \mrc{} $(\sigma,\ql,\qtau,\rho)$ if
\[
    VU\Pi V = \Pi V, VU\Pi \ql \Pi V = \Pi \ql \Pi V \text{ and } VU\Pi \rho = \Pi \rho,
\]
where $\Pi$ is the ergodic projection of $\qtau$, and $U$ is some distributor
for $V$.
\end{definition}

We now explain what this weak bisimulation means stochastically. It is a known result from perturbation theory that $P_\tau(t) =e^{Qt} = e^{(\ql + \tau\qtau)t}$ (uniformly) converges to $\Pi e^{\Pi\ql\Pi t}$ when $\tau\rightarrow\infty$, where $\Pi$ is the ergodic projection at zero of $\qtau$. The matrix $\Pi e^{\Pi\ql\Pi t}$ is a \emph{non-standard} transition
matrix~\cite{chung} being discontinuous at $t=0$. It is, however, a transition
matrix of a \emph{discontinuous} \mc{}~\cite{doeblin}, a stochastic process that generalizes
standard \mcs{} by being allowed to perform infinitely many transitions in
finite time. Strong bisimulation easily extends to discontinuous \mrcs{} by
adding the condition $VU\Pi V = \Pi V$ to Definition~\ref{def::str-bsm-mrc}.
Weak bisimulation can then be interpreted as a strong bisimulation on the
discontinuous \mrc{} that is obtained when $\tau\rightarrow\infty$. The idea of
taking the limit is to treat the transitions from $\qtau$ as instantaneous
whenever we abstract from them; if a transition takes time it must be
considered observable. This exactly was our motivation
in~\cite{QEST2006,TrckaPhD} where we defined the notion of $\tau$-lumpability,
and it explains why the conditions for weak bisimulation exactly match the
conditions of $\tau$-lumpability.

The concept of weak lumping, i.e.\ of reduction modulo weak bisimulation, can
also be introduced here. However, while we were able to use $\transp{V}$ for
transition systems as a special distributor ensuring that the order of
application of $\tau$-closure and lumping is irrelevant, here a more
complicated analysis is needed (in the real-number matrix theory, $\transp{V}$
is not even a distributor!). One of the main results
from~\cite{QEST2006,TrckaPhD} is the notion of a $\tau$-distributor $W$ that is
used to define the lumped process as $(\sigma V, W\ql V, W\qtau V, W\rho)$.
This is a special distributor that gives a lumped chain of which the limit is
the lumped version of the limit of the original chain. In other words, it
ensures that the following diagram commutes:
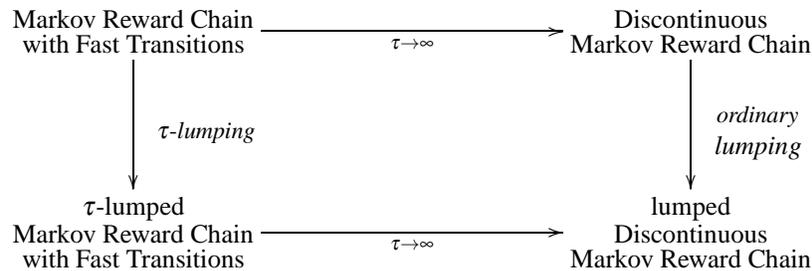
\begin{displaymath}\small
 \xymatrix@R1.7cm@C4cm{
 \txt{Markov Reward Chain\\ with Fast Transitions} \ar[r]_{\tau\rightarrow\infty} \ar[d]^{
 \begin{tabular}{c}\footnotesize$\tau$-\emph{lumping}\end{tabular}} &
 \txt{Discontinuous\\ Markov Reward Chain} \ar[d]^{\begin{tabular}{c}\footnotesize\emph{ordinary}\\\emph{lumping}\end{tabular}} \\
\txt{$\tau$-lumped\\ Markov Reward Chain\\ with Fast Transitions} \ar[r]_{\tau\rightarrow\infty} & \txt{lumped\\
        Discontinuous\\ Markov Reward Chain}}
\end{displaymath}
The precise definition of $W$ is complicated and outside the scope of this
paper; it can be found in~\cite{TrckaPhD} together with the proof of the above
diagram. Note that although a $\tau$-distributor $W$ and the special
distributor $\transp{V}$ from the transition system setting appear to have no
connections at all, they actually represent the same thing. In~\cite{TrckaPhD}
we have shown that $W$ is a distributor that satisfies $\Pi V W \Pi$,
and is such that the ergodic projection of $W\qtau V$ is $W\Pi V$. Recall that
these are exactly the  properties of $\transp{V}$ that we established in
Theorem~\ref{thm::pivw-lts}, interpreted in the boolean matrix setting.

Having defined the weakly lumped process, we can speak of properties that are
preserved by lumping. It can, e.g., be shown that the expected reward at $t$ is
the same for the two chains in the limiting case of $\tau$.

To define branching bisimulation for \mrcs{} we first note that $\qtau \sqcap
V\transp{V}$ is in general not a generator matrix; the diagonal of $\qtau$ is
never affected by this operation. We will, however, conveniently assume that
the obvious small adaptation has been made on the diagonal of $\qtau$ to turn $\qtau \sqcap V\transp{V}$ into a generator, and we will denote the obtained matrix $\qtauV$.

Putting $(\sigma,\ql,\qtau,\rho)$ instead of $(\sigma,A,S,\rho)$ in
Definition~\ref{def::branch-bsm-lts} and using the cancelation property valid
in $\mspc{n}{n}$, we obtain the following definition of branching bisimulation
for \mrcs{}.

\begin{definition}[Branching bisimulation-MRC]\label{def::branch-bsm-mrc}
A collector matrix $V\in\zo^{n\times N}$ is called a \emph{branching bisimulation
on} the \mrc{} $(\sigma,\ql,\qtau,\rho)$ if
\begin{equation*}
    VU\PiV \qtau V = \Pi_V \qtau V,\quad VU\PiV \ql V = \PiV \ql V, \quad \text{and} \quad VU\PiV \rho = \PiV \rho,
\end{equation*}
where $\PiV$ is the ergodic projection of $\qtauV$, and $U$ is some distributor
for $V$.
\end{definition}

Note that there is actually no branching structure to be preserved in \mrcs{}
as (almost) every $\tau$, being instantaneous in the limit, would have priority over any
regular (exponential) transition. This makes the usefulness of branching
bisimulation in this setting questionable. Moreover, in contrast to transition
systems theory, we have not been able to prove that the above definition ensures that every branching bisimulation is also a weak bisimulation, nor whether there exists a commuting diagram similar to the one for strong and weak lumping.

\section{Conclusions and Future Work}
We used  matrix theory as a unified framework to present the theory of
transition systems and Markov reward chains, and of their bisimulations. The
notions of strong, weak, and branching bisimulation on transition systems were
first coded in terms  of matrix equalities. The compactness and the algebraic
power of this representation is then illustrated in few important theorems. The
same matrix equalities were next interpreted in the Markov reward chain
setting, directly leading to the notions of strong, weak, and branching
bisimulation there. The obtained notion of strong and weak bisimulation were
shown to coincide with the existing notions of ordinary and $\tau$-lumpability
respectively. The obtained notion of branching bisimulation is new, but its
properties are unknown and its usefulness is yet to be seen.

In \cite{TrckaPhD,MSTdeVPEVA09} another form of aggregation in \mrcs{} is
presented. In contrast to
 lumping this method  always
eliminates all internal steps by allowing states to be split into multiple
classes. For future work we schedule to investigate whether this new reduction
 would lead to an interesting notion in the transition system setting. We also
plan to see if the bisimulation-up-to technique~\cite{sangiorgi} (formalized in
matrix terms in~\cite{TrckaPhD}) is applicable to \mcs{}.

\bibliographystyle{plain}
\def\sortunder#1{}
\bibliography{bibliography}

\end{document}